\documentstyle[eqsecnum,aps]{revtex}
\begin{document}
%\draft
\title
{.\\
Non-occurrence of Trapped Surfaces and  Black Holes in
Spherical Gravitational Collapse: An Abridged Version}

\author{Abhas Mitra}
\address{Theoretical Physics Division, Bhabha Atomic Research Center,\\
Mumbai-400085, India\\ E-mail: amitra@apsara.barc.ernet.in}

%\date{\today}

\maketitle
\vskip 1.5cm
%\begin{abstract}
By using the most general form of Einstein equations for General
Relativistic (GTR) spherical collapse of {\em an isolated} fluid having
arbitrary equation of state  and radiation transport properties, we show
that they obey a {\em Global Constraint}, $2GM(r,t) /R(r,t) c^2 \le 1$,
where $R$ is the ``invariant circumference radius'', $t$ is the comoving
time, and $M(r,t)$ is the gravitational mass enclosed within a comoving
shell $r$.
 This inequality specifically shows that, contrary to the
traditional intuitive Newtonian idea, which equates the total
gravitational mass ($M_b$) with the fixed baryonic mass ($M_0$), the {\em
trapped surfaces} are not allowed in general theory of relativity (GTR),
and therefore, for continued collapse, the final gravitational mass $M_f
\rightarrow 0$ as $R\rightarrow 0$.  This result  should be valid for all
spherical collapse scenarios including that of collapse of a spherical
homogeneous dust as enunciated by Oppenheimer and Snyder (OS).  Since the
{\em argument of a logarithmic function cannot be negative}, the Eq. (36)
of the O-S paper ($T\sim
\ln{y_b+1\over y_b-1}$) categorically demands that $y_b=R_b/ R_{gb} \ge
1$, or $2G M_b/R_b c^2
\le 1$, where $R_b$ referes to the invariant radius at the outer boundary.
Unfortunately, OS worked with an approximate
 form of Eq. (36) [Eq. 37], where this fundamental constraint
got obfuscated. And although OS noted that for a
finite value of $M(r,t)$ the spatial metric coefficient for an internal
point {\bf fails to blow up even when the collapse is
complete}, $e^{\lambda(r<r_b)} \neq \infty$ for $R\rightarrow 0$, they,
nevertheless, {\bf ignored} it,
and, failed to realize that such a problem was occurring because they were
assuming a finite value of $M_f$, where $M_f$ is the value of the finite
gravitational mass, in violation of
their Eq. (36).

Additionally, {\em irrespective of the gravitational collapse
problem}, by analyzing the properties of the Kruskal
transformations we show that in order that the actual radial
geodesics remain {\em timelike}, finite mass Schwarzschild Black
Holes cannot exist at all.

 Our work shows that as one attempts to arrive at the singularity,
$R\rightarrow 0$, {\bf the proper radial length} $l = \int
\sqrt{-g_{rr}}~dr \rightarrow \infty$ (even though $r$ and $R$ are
finite), {\em and the collapse process continues indefinitely. During this
indefinite journey, naturally, the system radiates out all available
energy}, $Q\rightarrow M_i c^2$, because trapped surfaces are not formed.
And this categorically shows that GTR is not only ``the most beautiful
physical theory'', but also, is the only, naturally, singularity free
theory (atleast for isolated bodies), as intended by its founder,
Einstein. However, this derivation need not rule out the initial
singularity of ``big bang'' cosmology because the universe may not be
treated as an ``isolated body''.

There is a widespread misconception, that recent astrophysical
observations have proved the existence of Black Holes.  Actually, observations
suggest existence of compact objects having masses greater than the upper
limit of {\em static} Neutron Stars. The present work also allows to have
such massive compact objects. It is also argued that there is evidence
that part of the  mass-energy accreting onto several stellar mass (binary)
compact objects or massive Active Galactic Nuclei is getting ``lost'',
indicating the presence of an Event Horizon. Since, we are showing here
that the collapse process continues indefinitely with local 3-speed $V
\rightarrow c$, accretion onto such Eternally Collapsing Objects (ECO) may
generate little collisional energy out put. But, in the frame work of
existence of {\em static} central compact objects, this small output of
accretion energy would be misinterpreted as an ``evidence'' for Event
Horizons. Thus the supposed BHs are actually massive compact ECOs.

%\end{abstract}
\vskip 1cm

\newpage

\section{Introduction}
One of the oldest and most fundamental problems of physics and
astrophysics is that of gravitational collapse, and, specifically, that of
the ultimate fate of a sufficiently massive collapsing body\cite{1,2}.
Most of the astrophysical objects that we know of, viz. galaxies, stars,
White Dwarfs (WD), Neutron Stars (NS),  in a broad sense, result from
gravitational collapse. And in the context of classical General Theory of
Relativity (GTR), it is believed that the ultimate fate of sufficiently
massive bodies is collapse to a Black Hole (BH)\cite{3}. A spherical
chargeless BH of (gravitational) mass $M_b$ is supposed to occupy a region
of spacetime which is separated by a hypothetical one-way membrane of
``radius'' $R_{gb} = 2G M_b/c^2$, where $G$ is the Newtonian gravitational
constant and $c$ is the speed of light. This membrane, called, an event
horizon, where (local)
acceleration due to gravity blows up,
is supposed to contain a central singularity at $R=0$, where the other
 physically relevant quantities like (local) energy density, (local) tidal acceleration, and components of
the Rimmenian curvature tensor diverge. However, although such ideas are,
now, commonly believed to be elements of ultimate truth, the fact remains
that, so far, it has not been possible to obtain any analytical solution
of GTR collapse equations for a physical fluid endowed with pressure
($p$), temperature $(T)$ and an equation of state (EOS). And the only
situation when these equations have been solved (almost) exactly, is by
setting $p\equiv 0$, and further by neglecting any density gradient, i.e.,
by considering $\rho = constant$\cite{4}. It is believed that these
(exact) asymptotic solutions actually showed the formation of BH in a
finite comoving proper time $\tau_{gb}$. However, this, assumption of
perfect homogeneity is a very special case, and, the speed of sound $c_s
=(dp/d\rho)^{1/2}=\infty$ in such a case if $p>0$. And this is not allowed
by GTR or Special Theory of Relativity (STR). In fact, now many authors believe
that for a more realistic inhomogeneous dust, the results of collapse may
be qualitatively different\cite{5,6,7}. These authors, on the strength of
their semi-analytical and numerical computations, claim that the resultant
singularity could be a ``naked'' one i.e., one for which there is no
``event horizon'' or atleast some light rays can escape the singularity.
There has also been some effort to study the final stages of collapse by
assuming the presence of positive and likely negative pressure gradients
subject to the inherent difficultities and limitations of any ``direct approach''.
For instance, recently Cooperstock et al.\cite{8} undertook such a study
and tentatively concluded that for positive pessure gradients a BH is
likely whereas for occurrence of negative pressure gradients a ``naked
singularity'' may form.
Therefore light may emanate from a naked singularity and reach a distant
observer. A naked singularity may also spew out matter apart from light
much like the White Holes.  In other words, unlike  BHs, the naked
singularities are visible to a distant observer and, if they exist, are of
potential astrophysical importance. However, according to a celebrated
postulate by Penrose\cite{9}, called ``Cosmic Censorship Conjecture'', for
all realistic gravitational collapse, the resultant singularity must be
covered by an event horizon, i.e, it must be a BH. And many authors
believe that the instances of occurrences of ``naked singularities'' are
due to fine tuned artificial choice of initial conditions or because of
inappropriate handling and interpretation of the semi-analytical
treatments. In this paper, we are not interested in such issues and would
avoid presenting and details about the variants of naked singularities
(strong, weak, local, global, etc.) or the variants of the censorship
conjecture.

\subsection {Kelvin - Helmholtz (KH) Process}
As a self-gravitating body contracts, it radiates and the same time its
internal energy increases. And the internal energy can have two
contributions:
\begin{equation}
E_{in} = E_{T} + E_{cold}
\end{equation}
where $E_{T}=\int e_{T} d\cal{V}$ is the temperature
dependent thermal part of the
internal energy and $E_{cold}= \int e_{cold} d\cal{V}$ is due to the
 pure degeneracy effects and which may
exist in certain cases even if the star is assumed to be at a temperature
$T =0$. Here $d\cal{V}$ is an element of proper volume.
The corresponding energy densities are
\begin{equation}
e_T = {(3/\pi^2)^{1/3} mc^2\over 6 (\hbar c)^2} n^{1/3} T^2; \qquad
T\rightarrow 0
\end{equation}
Here $m$ is nucleon mass and $n$ is nucleon number density.  Actually,
when the body is really degenerate, this kind of splitting of $E_{in}$ can
be done only in an approximate manner. For example, if it is assumed that
a degenerate ideal neutron gas is close to $T=0$, i.e, $ T \ll T_{Fermi}$,
then one may approximately take the first term  (lowest order in $T$) of
an infinite series to write the above expression.  On the other hand, if
the temperature is indeed much higher than the corresponding Fermi
temperature, degeneracy will vanish, $e_{cold}
\rightarrow 0$, and the entire energy density will be given by the thermal
contribution:
\begin{equation}
e= e_{T} = {3\over 2} n kT; \qquad T\rightarrow \infty
\end{equation}
where $k$ is the Boltzman constant, and
\begin{equation}
e_{cold}= {2\over 3} p_{cold} = {2 (3\pi^2)^{2/3}\over 15} {\hbar^2\over
m} n^{5/3}; \qquad \gamma_t =5/3
\end{equation}
where $\gamma_t$ is the effective ratio of specific heats.
Since it is not known beforehand, how $T$ would evolve, in principle, one
should work with an expression for $e_T$ (an infinite series) valid for
arbitrary $T$. But, it  is not possible to do so even for an ideal Fermi
gas. As to the actual EOS of nuclear matter at a finite $T$, it may be
remembered that, it is an active field of research and still at its
infancy. Thus, in practice it is impossible to make much headway without
making a number of simplifying assumptions because of our inability to
{\em self consistently} handle: (i) the equation of state (EOS) of matter
at arbitrarily high density and temperature, (ii) the opacity of nuclear
matter at such likely unknown extreme conditions, (iii) the associated
radiation transfer problem and all other highly nonlinear and coupled
partial differential (GTR) equations (see later).

One may start the numerical computation by presuming that indeed the
energy liberated in the process $Q \ll M_i c^2$, i.e., the effect of GTR
is at best modest. Then, it would naturally be found that the temperature
rise is moderate and depending on the finite grid sizes used in the
analysis and limitation of the computing machine, one may conclude that
the formalism adopted is really satisfying, and then find that $Q \ll M_i
c^2$\cite{10,11}.  Meanwhile, one has to extend the  presently known
(cold) nuclear EOS at much higher densities and maintain the assumption
that the rise in temperature is moderate. Because if $T$ is indeed high,
in the diffusion limit, the emitted energy $Q \sim T^4$ would be very
high, and the value of $M_f$ could drop to an alarmingly low value. Thus,
for the external spacetime, one needs to consider the Vaidya metric\cite{12,13,14}.
Actually, even when, $T$ is low, it is extremely difficult to self
consistently handle the coupled energy transport problem.

It may appear that, the practical difficulties associated with the study
of collapse involving densities much higher than the nuclear density can
be avoided if one starts with a very high value of $M_i$, say, $10^{10}
M_\odot$ (solar mass). Then {\em if one retains the assumption} that $M_i = M_f$, one
would conclude that an ``event horizon'' is formed at a density of $\sim
10^{-4}$ g cm$^{-3}$ for which the EOS of matter is perfectly well known:
\begin{equation}
\rho_g = {3c^6\over 32 \pi G^3 M_i^2} \approx 2 \times 10^{16} {\rm g~cm}^{-3}~ \left({M_i\over
M_{\odot}}\right)^{-2}
\end{equation}
Here $\rho_g$ is the closure density. It may be reminded that this
$\rho_g$ or the generic mass-energy density in GTR $\rho$ {\em does not} include
the (negative) contribution of self-gravity. On the other hand, $\rho$ includes rest mass
and all internal energy densities. It is well known that, in GTR, there is
no locally defined gravitational energy density; however the effect of
(negative) self-gravity appears in {\em globally} defined concepts like
enclosed Schwarzschild (gravitational) mass:
\begin{equation}
M=\int \rho dV
\end{equation}
Here $dV$ is an element of ``coordinate volume'' element and not the
locally measured physically meaningful ``proper volume'' element $d{\cal
V}$. While $\rho$ refers only to the material energy density and does not
take into account the contribution to the energy from gravity and while
$dV$ is the coordinate volume and not the physical proper volume, $d{\cal
V}$, the combination of these elements in Eq. (1.6) work together to yield
the correct total energy of the body {\em (including) the contribution
from gravity}\cite{15}.  {\em What is overlooked}
in the traditional
interpretation of Eq. (1.5) is that, this  expression is {\bf incorrect}, and the
correct expression should involve $M_f$ and not $M_i$:
\begin{equation}
\rho_g = {3c^6\over 32 \pi G^3 M_f^2} \approx 2 \times 10^{16} {\rm g~cm}^{-3}~ \left({M_f\over
M_{\odot}}\right)^{-2}
\end{equation}
Once we are
assuming that an event horizon is about to form, {\em we are endorsing the fact
that we are in the regime of extremely strong gravity}, and, therefore for
all the quantities involved in the problem,  a real GTR estimate has to be
made without making any prior Newtonian approximation.

  To further
appreciate this important but conveniently overlooked point (by the
numerical relativists), note that, the strength of the gravity may be
approximately indicated by the ``surface redshift'', $z_s$, of the
collapsing object, and while a Supermassive Star may have an initial value
of $z_s$ as small as $10^{-10}$, a canonical NS has $z_s \sim 0.1$, while
the Event Horizon, irrespective of the initial conditions of the collapse,
has got $z_s =\infty$! Therefore all Newtonian or Post Newtonian estimates
or the conclusions based on such estimates have {\em little relevance for
actual gravitational collapse problem}.

 As a result, the integrated value of $Q$
may tend to increase drastically, and this would {\em pull down the
running value of} $M_f = M_0 - Q/c^2$ and $R_{gb}$ to an alarming level!
 At the same time, of
course, the value of $R$ is decreasing. But how would the value of
$R_b/R_{gb}$ would evolve in this limit? Unfortunately, nobody has ever,
atleast in the published literature, tried to look at the problem in the
way it has been unfolded above. On the other hand, in Newtonian notion,
the value of $M_f$ is permanently pegged at $M_0$ because energy has no
mass-equivalence (although in the corpuscular theory of light this is not
so, but then nobody dragged the physics to the $R\rightarrow R_g$ limit
seriously then).  So, in Newtonian physics\cite{1,2}, or in the intuitive thinking
process of even the GTR experts\cite{3}, the value of
\begin{equation}
{2 G M \over R c^2} \equiv {2G M_0 \over R c^2} \rightarrow \infty;
\qquad R\rightarrow 0
\end{equation}
 and the idea of a trapped surface seems to be most natural.  But, in GTR,
we cannot say so with absolute confidence even if we start with an
arbitrary high value of $M_i$ because, in the immediate vicinity of $R
\rightarrow R_{g}$, the running value of $M$ may decrease in a fashion
which we are not able to fathom either by our crude qualitative arguments,
based on GTR, or by numerical computations plagued with uncertain physics
and inevitable machine limitations.
And, if $M_f$ drops to an alarming level, the actual value of $\rho_g$
can rise to very high values. Thus all the
 difficulties associated with the numerical study of  the collapse of a stellar
mass object
 may reappear  for any value of $M_0$  unless one hides the
nuances of GTR and other
detailed physics with favorable and simplifying assumptions and approximations.
To seek a real answer for such questions,
we need to handle GTR carefully and exactly in a manner different from
this qualitative approach.

\section {Spherically Symmetric Gravitational Field}
The most general form of a spherically symmetric metric, after appropriate
coordinate transformations, can be brought to a specific Gaussian form \cite{15,16,17,18},
:
\begin{equation}
ds^2 =   A^2(x_1,x_0) dx_0^2 -  B^2(x_0,x_1) dx_1^2 -  R^2 (d\theta^2 +\sin^2 \theta d\phi^2)
\end{equation}
where $ A$ and $ B$  are to be determined self consistently for a given
problem. For the simplification of computation, it is customary to express:
\begin{equation}
 A^2 = e^\nu; \qquad  B^2 =e^\lambda
\end{equation}
Here $x_1$ is an appropriate radial marker (coordinate),  and $x_0$ is the
coordinate time.
For the spherical metric, if we consider a $x_0 = constant$
hypersurface and pick up a curve (circle) with $x_1=constant$ and
$\theta=\pi/2$, the value of the invariant line element would be
\begin{equation}
 ds = R d\phi
\end{equation}
The invariant circumference of the $x_0=constant$ circle would be
$2 \pi R$, and thus, we identify $R$ as the (invariant)
circumference variable. The $R$ thus {\em defined acquires a
physical significance and appears to be related to the luminosity distance
 for astronomical observations}. Clearly this property of $R$ as
the invariant circumference radius is a pure consequence of spherical
symmetry irrespective of whether there is a probable coordinate
singularity or not.
One has a Schwarzschild
coordinate system when one chooses $x_1 =R$. If we label the
corresponding time as $T$, we have
\begin{equation}
ds^2 = e^\nu dT^2 - e^\lambda dR^2 - R^2 (d\theta^2 +\sin^2 \theta d\phi^2)
\end{equation}
 And if we demand that $e^\nu \rightarrow 1$ as $R\rightarrow \infty$, we
can identify $T$ as the proper time of a distant observer.

\subsection {Exterior Schwarzschild Metric (ESM)}
But, now, suppose we are going to describe a truly ``vacuum'' exterior
spherical solution, an exterior spacetime region not containing a
single ``particle'' or photon.
The actual solution for the vacuum exterior spacetime region
 was found by Schwarzschild in 1916\cite{19}:
\begin{equation}
ds^2 = g_{TT} dT^2 + g_{RR} dR^2 - R^2 (d\theta^2 +\sin^2 \theta d\phi^2)
\end{equation}
with
\begin{equation}
g_{TT} = \left(1 -{R_{gb}\over R}\right); \qquad g_{RR}= - \left(1-{R_{gb}
\over R}\right)^{-1}; \qquad R_{gb}= {2GM_b\over  c^2}
\end{equation}
where $M_b$ is the total gravitational mass of the system.
The time parameter $T$ appearing here  is {\em not}
any comoving time measured by a clock attached to the test particle. On
the other hand, as mentioned above,
 $T$ acquires a distinct physical meaning as the {\em
 proper time measured by a distant inertial observer} $S_\infty$.
 Thus both $R$
and $T$ have some sort of absolute meaning in the External
Schwarzschild Metric (ESM). The singularity of this {\em vacuum}
mertic at $R=R_{gb}$ is obvious, and, as per Landau and Lifshitz\cite{15}, if any fluid is
squeezed to $R\le R_{gb}$, this singularity means that, the body
cannot remain static in such a case. On the other hand, the body
must be collapsing, if it ever reaches the Schwarzschild surface.
If during the preceding collapse process, the body radiates, the
vacuum condition would break down and the metric would simply
become inappropriate to describe a radiating scenario. However,
if the fluid is a dust, the collapse process would be radiation
free, and the vacuum solution must continue to hold good exterior
to the collapsing dust at every stage because the derivation of
Eq. (2.6) is absolutely general (except of course for the
``vacuum'' condition).

In contrast, a comoving frame (COF), by definition, can be
constructed in a region {\em filled with mass-energy} and can be
naturally defined in the interior of any fluid. And since in this
case $x_1$ is fixed with the particle and time is measured by the
clock moving with the test particle, the question of any
coordinate singularity  does not arise.

To appreciate, again, the point that $R$ always retains it physical significance as
the ``metric distance'' for spherical symmetry, consider the case of the
infall of a free particle towards the central singularity of a supposed
massive BH (if it would exist). The conventional wisdom is that when the
particle crosses the Event Horizon, $R$ and $T$ would exchange their
roles, i.e,  $R$ would be time like while $T$ would be space like.
However, note that, even in this case {\em the ``location'' of the  central
singularity is still denoted by} $R=0$ and {\em not by} $T=0$ or any
$T=T_0$. Further, if one would like to introduce a totally different
coordinate system, such as Kruskal Coordinates $u$ and $v$ (see later),
the central singularity is still expressed, most conveniently,  as
$R=0$ and not as $u=0$ or $v=0$. Even if one would mechanically express
the central singularity as, say, $u=u_0$, even then, the idea of $R=0$ stalks
in the back ground, and $u$ needs to be calibrated against $R$ for a
meaningful physical description. Without the support of the concept of $R$, the
essential spatial distinction between two events cannot at all be described.

\section{Formulation of the Collapse Problem}
The general
formulation for the GTR collapse of a perfect fluid, by ignoring any
emission of radiation, i.e, for adiabatic collapse, is well developed\cite{18,20,21}.
 Although, our central result would not depend on the details of the
numerous equations involved in the GTR collapse problem, yet, for the sake
of better appreciation by the reader, we shall outline the general
formulation of the GTR spherical collapse problem, and refer, the reader
to the respective original papers for greater detail.
It is most appropriate and simple to formulate the collapse problem
in the comoving frame $x_0 =t$ and $x_1=r$:
\begin{equation}
ds^2 =   A^2(r,t) dt^2 -  B^2(r,t) dr^2 -  R^2(r,t) (d\theta^2 +\sin^2
\theta d\phi^2)
\end{equation}
The most physically significant choice for $r$ is one which
corresponds to the fixed number of baryons inside a shell of
$r=r$. The element of proper time is naturally obtained by
following the particle $r=constant$\cite{15} in radial motion
\begin{equation}
ds =d\tau = A dt
\end{equation}
Now we set $c=1$. On the other hand, for this {\em static metric}, the element of proper
length along the radial motion of the fluid is\cite{15}
\begin{equation}
dl = B dr
\end{equation}
 At first sight,
one might think if $r$ is really a  ``comoving coordinate'' and $dr=0$
for a particular $r$,
then what does a derivative with respect to $r$ or a derivative for $r$ mean?
Here $dr$ is to be interpreted as the difference in the value of
$r$ between two close by shells having $r=r$ and $r=r+dr$, and
also the increment of a particular shell during the overall fluid
motion. To understand the latter interpretation of $dr$, we can first
consider a 
Newtonian hydrodynamics problem. Suppose we take snapshots of the fluid
shells corresponding to comoving times $t$ and $t+dt$ recorded by a given
clock fixed initially at $r=r$. Now, if we superimpose the two snapshots,
we can measure the the radial increment of a particular shell by comparing
its relative position 
against the
backdrop of  the other. The corresponding derivative $dr/dt$ will give the
local speed of the fluid in terms of comoving coordinates in the Newtonian case.
In the GTR case too, one can similarly define a derivative $dr/dt$ and use
it for further defining appropriate 3 -speed (see later).
Before we proceed, we may define few variables which, to start with may be
taken as pure symbols\cite{18,20,21,22,23}:
\begin{equation}
U\equiv {dR\over d\tau} ={dR\over A dt}\mid_{r=r} = {{\dot R}\over A}
\end{equation}
and
\begin{equation}
\Gamma \equiv  {dR\over dl}= {dR\over B dr}\mid_{t=t} = {R'\over B}
\end{equation}
where a prime denotes differentiation with respect to $r$ and an overdot
denotes the same with respect to $t$. Note that while $U$ has the nature
of a partial derivative with respect to $t$, it is a {\em total derivative} with
respect to $\tau$ because the concept of a fixed $r$ is inherent in the
concept of $d \tau$. Similarly, while $\Gamma$ has the nature of a partial
derivative with respect to $r$, it is a {\em total derivative} with respect to
$l$ because the concept of a fixed $t$ is inherent in the concept of
element of proper length $dl$ (see pp. 180-181 of ref. 22 and pp. 150-151
of ref. 23).

We now recall the Einstein
equation itself:
\begin{equation}
R_{ik} = 8\pi G \left( T_{ik} - {1\over 2} T\right)
\end{equation}
where the energy momentum stress tensor for a perfect fluid, in the COF, is
\begin{equation}
T^r_r =T^\theta_\theta=T^\phi_\phi =- p; \qquad T^0_0 =\rho;\qquad T=T^i_i
=\rho-3p
\end{equation}
Here $p$ is the isotropic pressure (in the proper frame)
and the total energy density of the fluid
in the same frame (excluding any contribution from global
self-gravitational energy) is
\begin{equation}
\rho =\rho_0 + e
\end{equation}
where $\rho_0 = m n$ is the proper density of the rest mass, $n$ is the
number density of the baryons in the same frame (one can add leptons too),
 and $e$ is the proper
internal energy density.
Here $R_{ik}$ is the contracted (fourth rank) Rimennian curvature tensor
$R^i_{jkl}$, and, is called the Ricci tensor. In terms of the Christoffel
symbols
\begin{equation}
\Gamma^i_{kl} = {1\over 2} g^{im} \left({\partial g_{mk}\over \partial
x^l} + {\partial g_{ml}\over \partial x^k} - {\partial g_{kl}\over
\partial x^m}\right)
\end{equation}
the components of the Ricci tensor are
\begin{equation}
R_{ik} = {\partial \Gamma^l_{ik}\over \partial x^l} - {\partial
\Gamma^l_{il}\over \partial x^k} + \Gamma^l_{ik} \Gamma^m_{lm} -
\Gamma^m_{il} \Gamma^l_{km}
\end{equation}
 One also requires to use the local energy momentum
conservation law:
\begin{equation}
T^i_k; k =0
\end{equation}
where a semicolon, ``;'', denotes covariant differentiation:
\begin{equation}
T^i_ k; l = {\partial T^i_k \over \partial x^l} -\Gamma^m_{kl} T^i_m +
\Gamma^i_{ml} T^m_k = 0
\end{equation}
 One has to supplement these equations
with the equation for continuity of baryon number :
\begin{equation}
(nu^i); i =0
\end{equation}
Now, after considerable algebra, in the COF,
the Einstein equations become\cite{21} :
\begin{equation}
(R^0_0):~~~~~~ 4\pi G \rho R^2 R' = {1\over 2} \left( R+ {R {\dot R}^2\over
A^2} -{R R'^2\over B^2}\right)'
\end{equation}
\begin{equation}
(R^r_r): ~~~~~ 4\pi G p R^2 {\dot R} = -{1\over 2} \left( R +{R {\dot R}^2\over
A^2} -{R R'^2\over B^2}\right)^.
\end{equation}
\begin{equation}
(R^{\theta}_{\theta}, R^{\phi}_{\phi}):~~~~ 4\pi G (\rho+p)R^3=\left(R+ {R{\dot R}^2\over
A^2}-{RR'^2\over B^2}\right)+{R^3\over AB} \left[\left({A'\over
B}\right)'-\left({{\dot B}^.\over A}\right)\right]
\end{equation}
\begin{equation}
(R^r_0, R^0_r): ~~~~~~ 0 = {A'{\dot R}\over A} + {{\dot B} R'\over B} -
{\dot R}'
\end{equation}
Further, if we define a new function
\begin{equation}
M(r,t)  = 4\pi \int^r_0 \rho R^2 R' dr
\end{equation}
the $R^0_0$- field Eq. (3.14) can be readily integrated to
\begin{equation}
R + {R {\dot R}^2 \over A^2} - {R R'^2\over B^2} = 2 GM
\end{equation}
where the constant of integration has been set to zero because of the
standard central boundary condition
$M(0,t) =0$.
Here the coordinate volume element of the fluid is $dV =4 \pi R^2 R'dr$.
If we move to the outermost boundary of the fluid situated at a fixed
$r=r_b$, and demand that the resultant solutions match with the
exterior Schwarzschild solution, then we would be able to identify
$M(r_b)$ as the function describing the total (gravitational) mass  of the
fluid {\em as measured by the distant inertial observer} $S_\infty$.  For
an interpretation of $M(r, t)$ for the interior regions, we first, recall
that the element of proper volume is
\begin{equation}
d {\cal V} = 4\pi R^2 dl= 4 \pi R^2 B dr = {dV \over \Gamma}
\end{equation}
so that
\begin{equation}
 M(r,t) =  \int^r_0 \rho dV  = \int^r_0 (\Gamma \rho) d {\cal V}
\end{equation}

And this suggests that $\Gamma \rho$ is the energy density measured by $S_\infty$.
We have already inferred that $M_b$ is the total mass energy of the fluid
as seen by $S_\infty$. Then for a self-consistent overall description,
 we can interpret that, in general, $M(r,t)$ is the mass-energy within
$r=r$ and as sensed by $S_\infty$\cite{20}.

Now  by using the definitions $U$ and $\Gamma$ from Eqs. (3.4) and (3.5)
 in Eq. (3.19), we find
that
\begin{equation}
\Gamma^2 = 1 + U^2 - {2GM\over R}
\end{equation}
 Further using a compact notation
\begin{equation}
D_{t}= {1\over A} {d\over dt}\mid_{r=r}; \qquad D_{\rm r}= {1\over
B}{d\over dr}\mid_{t=t}
\end{equation}
the major adiabatic collapse equations turn out to be\cite{20,21}:
\begin{equation}
D_{\rm r}M=4\pi R^2 \rho \Gamma
\end{equation}
\begin{equation}
D_{\rm t}M= -4\pi R^2 p U
\end{equation}
\begin{equation}
D_{\rm t} U=-{\Gamma\over \rho+p} \left({\partial p\over \partial R
}\right)_{\rm t} -{M+4 \pi R^3 p \over R^2}
\end{equation}
\begin{equation}
D_{\rm t} \Gamma=-{U \Gamma \over \rho +p}\left({\partial p \over \partial
R}\right)_{\rm t}
\end{equation}
An immediate consequence of the last equation is that, if we assume a
$p=0$ EOS, $\Gamma$ will be time independent
$\Gamma (r, t)= \Gamma (r)$,
and for a fixed comoving coordinate $r$, $\Gamma$ would be a constant.
Further, the Eq. (3.26) shows that for $p=0$, we also have
$M (r, t)= M (r) =constant$.

\subsection{Collapse of a Physical Fluid}
We would emphasize that, for
studying the collapse of a physical fluid, it is absolutely necessary to
incorporate the radiation transport aspect in an organic fashion.  The
collapse equations were generalized to incorporate the presence of
radiation by several authors\cite{24,25,26,27,28}. Following Misner\cite{24} and Vaidya\cite{25}, we will
first treat the radiation part of the stress energy tensor in the geometrical optics limit:
\begin{equation}
E^{ik}= q k^ik^k
\end{equation}
where $q$ is both the energy density and the radiation flux in the proper
frame and $k^i=(1;1,0,0)$ is a null geodesic vector so that $k^ik_i=0$.

All one has to do now is to repeat the exercises  for an adiabatic fluid
outlined above by replacing the pure matter part of energy momentum
tensor with the total one:
\begin{equation}
T^{ik} = (\rho +p) u^i u^k + p g^{ik} + q k^i k^k
\end{equation}
Then the new $T^0_0$ component of the field equation, upon integration,
yields the new mass function:
\begin{equation}
M(r,t) = \int^r_0 4 \pi R^2 dR ( \rho + q v +q) = \int^r_0 d{\cal V }
[\Gamma (\rho + q) +q U]
\end{equation}
Had we treated the radiation transport problem without assuming a
simplified form of $E_{ik}$ and, on the other hand, in a most general
manner, following Lindquist\cite{28}, we would have obtained:
\begin{equation}
M(r,t) =  \int^r_0 d{\cal V }
[\Gamma (\rho + J) +H U]
\end{equation}
where
\begin{equation}
J= E^{00} = E^{ik} u_i u_k =q = comoving~ energy~ density
\end{equation}
and
\begin{equation}
H = E^{0R} = average~ radial~ flux
\end{equation}
This definition of $M$ may be physically interpreted in the following way:
while $(\rho+q)$ is the locally measured energy density of matter and
radiation, $\Gamma(\rho+q)$ is the same sensed by $S_\infty$ ($\Gamma \le
1$). Here, the radiation part may be also explained in terms of ``
gravitational red -shift''.  And the term $HU$ may be interpreted as the
Doppler shifted flux seen by $S_\infty$\cite{24,25,28}.
Although, the collapse equations, in general will change for such a
general treatment of radiation transport, {\em  the generic constraint
equation involving} $\Gamma$ {\em incorporates  this new definition of}
$M$ and remains unchanged:
\begin{equation}
\Gamma^2 = 1 + U^2 - {2 G M\over R}
\end{equation}
In the following, we list the other major collapse equations for the
simplified form of $E^{ik}$ only:
\begin{equation}
D_{\rm r}M=4\pi R^2\left[\Gamma (\rho +q)+ U q\right]
\end{equation}
\begin{equation}
D_{\rm t}M =- 4 \pi R^2 p U- L (U+\Gamma)
\end{equation}
\begin{equation}
D_{\rm t}U=-{\Gamma \over \rho+p} \left({\partial p \over \partial
R}\right)_{\rm t} -{M+4\pi R^3 (p+q)\over R^2}
\end{equation}
\begin{equation}
D_{\rm t}\Gamma=-{U\Gamma\over \rho +p}\left({\partial p\over \partial
R}\right)_{\rm t} +{L\over R}
\end{equation}
where the comoving luminosity is
\begin{equation}
L=4\pi R^2 q
\end{equation}
Even if there is no question of a strict exact solution (numerical or
analytical) for such a fluid, it is believed by practically all the
authors that a physical fluid will necessarily collapse to a singularity
in a finite proper time; and the debate hinges on whether the singularity
would be a BH or a naked one.  The modern conviction in the inevitability
of the occurrence of spacetime singularities in a general gravitational
collapse of a sufficiently massive configuration stems on the strength of
singularity theorems\cite{3}. Probably, the first singularity theorem, in
the context of spherical collapse, was presented by Penrose\cite{29} where
it was explicitly shown that once a trapped surface is formed, $2G M(r,t)
/R >1$, the collapse to the central singularity is unavoidable.  Since
then many authors like Hawking, Geroch, Ellis, including Penrose himself,
have proposed various forms of singularity theorems\cite{3}.  In the next
section, we shall show that the most innocuous
assumption behind the singularity theorems, namely, the assumption that

(1)  the manifold should contain a trapped surface either in
the past or future,

{\bf is actually not obeyed by the collapse equations}.

It is clear that, {\em if we are able to show that trapped
surfaces are not formed even for the most idealized case of a nonrotating
perfectly spherical perfect fluid not having any resistive agent like a
strong magnetic field, certainly trapped surfaces would not form in more
complicated situations}.

However, even before we present our deivation, it may be pointed out that
we are aware of atleast one review article by Senovilla\cite{29} which
specifically describes the possibilty that the final state of a
gravitational collapse may be singularity free (see subsection 7.2 of this
reference).

\section {Physical Speed V}
It is of utmost importance to be able to properly define the quantity,
$V$, which is the speed of the test particle or the fluid element measured
by a given static observer in a certain coordinate system.
For a {\em static field} such as one considered here, the
physical velocity in a generic coordinate system is given by
using Eq.(88.10)  or the the preceding unnumbered lowermost equation of pp.
250 of Landau and Lifshitz\cite{15}. In particular for the radial case, where,
$V_\phi =V_\theta =0$, we have
\begin{equation}
{V} = {dl \over d\tau}= {\sqrt{-g_{11}} dx_1\over \sqrt{g_{00}} dx^0}
\end{equation}
Here $dx_1/dx_0$ is to be interpreted as the total or ``convective''
derivative, and is, in general, non-zero even if $x_1$ is a comoving
coordinate.  In fact, except for Landau \& Lifshitz, no other  standard
textbook on GTR seems to contain this discussion on the definition of $V$
in detail, and, many experts on GTR also seem to be confused about this
important aspect.  One must note that, it is this $ V$ defined by Landau\&
Lifshitz which {\bf appears in the Local Lorentz transformations}.

We have already discussed in Sec. III, the meaning of a dynamical
derivative $dr/dt$ in the COF. While in a Newtonian hydrodynamic problem,
$dr/dt$ defines the fluid speed, in the GTR case, following Landau \&
Lifshitz, the 3-speed will be
\begin{equation}
{V}  = {dl \over d\tau}={B dr\over Adt} ={\sqrt{-g_{rr}} dr\over
\sqrt{g_{00}} dt}
\end{equation}

It may be recalled again that while $U$($\Gamma$) has the nature of
a partial derivative with respect to $t$($r$), it has the nature of a
total derivative with respect to $\tau$($l$).  The reader can convince
himself about this by going through references \cite{22,23}. This is just
similar to the fact while the fluid 4-velocity $u^i=  {\partial
x^i\over A\partial t}={dx^i\over d\tau}$ has a partial derivative nature with respect to $t$
it is actually a total derivative of $x^i$ with respect to the proper time
along the worldline of the fluid\cite{18}.
Then we note that the previously defined $U$ and $\Gamma$ are interlinked as
\begin{equation}
U=\Gamma V
\end{equation}

Here it may be also mentioned that the 3-velocity for the same fluid
element may be different in different coordinate system. For instance, the
3-velocity measured in a Schwarzschild field would be
\begin{equation}
V_{Sch} = {e^{\lambda/2} dR\over e^{\nu/2} dT}
\end{equation}
 And, in general, $V_{Sch} \neq V$(comoving). In particular, if the
(vacuum external) Schwarzschild coordinates really, suffer from a coordinate singularity at
$R=R_g$, it is likely that $V_{Sch}=1$ at $R=R_g$ and one would have
$V_{Sch} >1$ for $R <R_g$ (in the presence of matter-energy vacuum
Schwarzschild singularity is irrelevant). Here one may argue that at the coordinate
singularity, the Schwarzschild coordinates lose their meaning, and any
result obtained by extending them further is devoid of physical meaning.
However, since the comoving coordinates are by definition singularity
free and one can always read off the time recorded by a comoving clock, we
do not expect $V \ge 1$ in any truly non-singular domain of spacetime.

\section {The Central Proof: Heart of this Paper}
  For purely radial motions, one may ignore the
angular part of the metric to write:
\begin{equation}
ds^2 = g_{00} dt^2 + g_{rr} dr^2
\end{equation}
Again, {\em by definition}, the worldlines of photons or material
particles are null or time like, i.e., $ds^2 \ge 0$, so that
\begin{equation}
g_{00}\left[1 +\left( g_{rr} dr^2\over g_{00} dt^2\right)\right] \ge 0
\end{equation}
 or,

 \begin{equation}
g_{00} (1 -V^2) \ge 0
\end{equation}
Since in the singularity free COF, $g_{00} \ge 0$,  we find that $\gamma^{-2}
= 1- V^2 \ge 0$, where $\gamma$ is the Lorentz factor of the fluid. This
simple result expresses the fundamental fact that the 3-speed of photons
or material particles cannot exceed the speed of light (as long as
$g_{00} \ge 0$).

Further, we found in the  Section III that there exists a
 quantity $\Gamma^2$
\begin{equation}
\Gamma^2 \equiv {1\over A^2} \left({\partial R\over \partial
r}\right)^2\equiv {1\over -g_{rr}} \left({\partial R\over \partial
r}\right)^2 \ge 0
\end{equation}
and which is positive definite if so is $A^2 = -g_{rr}$.

Now, by substituting the above relationship (4.3)
into the right hand side of another global constraint (3.22) or (3.34),
we find that
\begin{equation}
\Gamma^2 = 1 + \Gamma^2 V^2 - {2G M(r,t)\over R(r, t)}
\end{equation}
Now by transposing, we obtain
\begin{equation}
\Gamma^2 (1 -V^2) = 1- {2G M(r,t)\over R(r, t)}
\end{equation}
This equation may be rewritten as
\begin{equation}
{\Gamma^2\over \gamma^2} = 1-{2G M(r,t)\over R(r, t)}
\end{equation}
 and this beautiful equation may be termed as the ``master
equation'' for spherical gravitational evolution of a system of a
fixed number of baryons.

 Since both $\Gamma^2 $ and $\gamma^2$ are positive definite, we find that
 the left side of Eq. (5.7) is positive definite;
and so must be the right hand
side of the same equation. Thus we obtain the most fundamental constraint for the
GTR collapse (or expansion) problem, in an unbelieveably simple
manner, as
\begin{equation}
{2G M(r,t)\over R(r, t)} \le 1 ; \qquad {R_{g} \over R} \le 1
\end{equation}

This shows that {\bf trapped surfaces do not form}.

Recall that it is
believed that the energy of an isolated body cannot be negative\cite{31}. From
physical view point, a negative value of $M_b$ could imply repulsive
gravity and hence is not acceptable.
When we accept this theorem(s), we find that the fundamental constraint
demands that {\em if the collapse happens to proceed upto} $R\rightarrow
0$, i.e., upto the central singularity, we must have
\begin{equation}
M(r, t) \rightarrow 0; \qquad R\rightarrow 0
\end{equation}
Remember here that the quantity  $M_0 =m N$ (which is the baryonic mass of
the star, if there are no antibaryons) is conserved as $M_f \rightarrow 0$.
 Physically, the $M=0$ state may result when the
{\em negative gravitational energy} exactly cancels the internal energy, the
{\em baryonic mass energy} $M_0$ and any other energy, and  which
is possible in the limit $\rho \rightarrow \infty$ and $p\rightarrow
\infty$.

\subsection{Singularity in Comoving Coordinates?}
Although, {\em comoving coordinates}, by definition, do not involve any
singularity unlike external Schwarzschild coordinates, in a desperate
attempt to ignore this foregoing small derivation, some readers might
insist that there could be a coordinate singularity somewhere so that
 $g_{00}$ could be negative in Eq.(5.3). If so, $(1-V^2)$ would be
negative implying that $V \ge 1$! And it might appear that, in such a case
Eq. (5.6) would lead to
$2GM/R \ge 1$. First we want to emphasize that while the (External)
{\em vacuum} Schwarzschild coordinate system might display such an
anomaly, i.e, the 3-speed of a free falling particle, measured in terms of
$R$and $T$ would indeed appear to exceed the speed of light once it is
inside the Event Horizon of a finite mass BH (if it would be allowed by GTR),
the {\em matter filled internal}
region would not display any coordinate singularity because the vacuum Sc.
metric coefficients are completely irrelevant there. In fact the entire
paper of Oppenheimer and Snyder\cite{4} is aimed at finding the metric
coefficients for the {\em matter filled internal} Schwarzschild coordinate
system all the way upto $R=0$, inside the supposed event horizon $R< R_g$.
In particular, as emphasized again and again, the comoving coordinates are
by definition singularity free. Even if one accepts
this incorrect possibility for a moment, our eventual result survives such
incorrect thinking in the following way.

Since, the determinant of the metric, $g = R^4 \sin^2 \theta ~g_{00}
~g_{rr}$ is always negative\cite{15} in all cases of coordinate singularity,
when, $g_{00} \le 0$, we would have $g_{rr} \ge 0$ and $A^2 \le 0$. Then it follows from
Eq.(5.3) that $\Gamma^2 <0$ so that the L.H.S. of Eq.(5.6) {\em
would be again positive}.
 Hence the
R.H.S. of the same too must be positive. And we get back Eq.(5.8), and thus,
it follows, {\em in a most general fashion}, that
\begin{equation}
{2 G M(r,t) \over R } \le 1
\end{equation}

Hence, even if we, for the sake of arguments, accommodate the possibility
(though quite incorrect) that the comoving coordinates might develop a
coordinate singularity, our eventual physical result that trapped surfaces
do not form remain unchanged.

\subsection{Previous Hints for M=0 Result}
While considering, the purely static GTR equilibrium configurations of dust,
Harrison et al.\cite{32}
discussed long ago that spherical gravitational collapse should
come to a {\em decisive end} with $M_f=M^*=0$, and, in fact, this
understanding was formulated as a ``Theorem''

``THEOREM 23. {\em Provided that matter does not undergo collapse at the
microscopic level at any stage of compression, then, -regardless of all
features of the equation of state - there exists for each fixed number of
baryons A a ``gravitationally collapsed configuration'', in which the
mass-energy $M^*$ as sensed externally is zero}.'' (Emp. by author).

In a somewhat more realistic way   Zeldovich and Novikov\cite{33}  discussed the
possibility of having an ultracompact configuration of degenerate fermions
obeying the EOS $p=e/3$ with $M\rightarrow 0$
and mentioned  the possibility of {\em having a machine for
which} $Q\rightarrow M_i c^2$.

  It is widely believed that Chandrasekhar's
discovery that White Dwarfs (WD) can have a maximum mass set the stage for
having a gravitational singular state with finite mass. The hydrostatic
equilibrium of WDs can be approximately described by Newtonian
polytropes\cite{34} for which one has $R\propto \rho_c^{(1-n)/2n}$, where
$\rho_c$ is the central density of the polytrope having an index $n$.  It
shows that, for a singular state i.e., for $\rho_c\rightarrow \infty$, one
must have $R\rightarrow 0$ for $n>1$; and Chandrasekhar's limiting WD
indeed has a {\em zero radius}\cite{34}. On the other hand, the mass of the
configuration $M\propto \rho_c^{(3-n)/2n}$.  And unless $n=3$,
$M\rightarrow \infty$ for the singular state. One obtains such a result
for Newtonian polytropes because they are really not meant to handle real
gravitational singularities. Fortunately, in the low density regime, when
the baryons are nonrelativistic and only electrons are ultrarelativistic,
the EOS is $p\rightarrow e/3$ and the
corresponding $n\rightarrow 3$. Then one obtains a finite value of $M_{ch}$
- the Chandrasekhar mass.

Now when we apply theory of polytropes for a case where the pressure is
supplied by the baryons and not only by electrons, we must consider GTR
polytropes of Tooper\cite{35}. It can be easily verified from Eq.  (2.24) of
this paper\cite{35}  that in the limit $\rho_c
\rightarrow \infty$, the scale size of GTR polytropes $A^{-1} \rightarrow 0$.
Further Eqs. (2.15) and (4.7) of the same paper  tell that $M \propto
K^{n/2}
\propto \rho_c^{-1/2} \rightarrow 0$ for $\rho_c\rightarrow \infty$. Thus,
a {\em proper GTR extension of Chandrasekhar's work would not lead to a BH
of finite mass}, but, on the other hand, to a singular state with
$M\rightarrow 0$.

In a different context, it has been argued that {\em naked singularities}
produced in spherical collapse must have $M_f=0$\cite{36}.

\section{Probable Regimes of Confusion}
Although, the Positive Energy Theorems\cite{31} probe whether
$M_b$ cannot only be zero but even negative, and although many
of the so-called naked singularity solutions correspond to zero
gravitational mass\cite{36}, we can foresee that many readers
would have difficulty in accepting our result.
 And,
instinctively, there may be a tendency to reject this
work on the basis of tangential and vague reasons. And though, we have
taken great care in developing several ideas, in the face
of the likely strong revulsion, several genuine or apparent confusions may
creep up:

\subsection{Baryonic and Gravitational Mass}
For some readers it might appear that a $M=0$ state corresponds to zero
baryon number $N=0$. Very clearly,  the
reader, because of instinctive Newtonian notion, in such a case would
incorrectly equate the gravitational mass with the baryonic mass: $ M \equiv M_0 = m
N$ (incorrect).

On the other hand, the gravitational mass of an isolated system
is just the aggregate of all kinds of energy associated with it, and for
any bound system, necessarily $M < M_0$. In particular, for the sake of
illustration, we may recall that in the weak gravity regime, we would have
\begin{equation}
M = M_0 + E_g + E_{in} + E_{kinetic}
\end{equation}

Here the gravitational energy term is always negative (even if $ M <0$)
and nonlinear. In the weak gravity regime it is $\sim -G M^2 /R$,
and as collapse proceeds, the grip of gravity
becomes tighter, and this is effected by the non linear nature of $E_g$.
As a result, the value of $M$, in general, steadily decreases in any
gravitational collapse, and, it is a
natural consequence that if we have a continued collapse, the value of
$M$  will hurtle downward and the system would try to seek a state of
``lowest energy''. In GTR, i.e., in Nature, the lowest energy corresponds
to $M=0$ and {\em not} to its Newtonian counterpart $E_N =E_g + E_{in} + E_{kinetic}=0$ (incorrect). Thus,
if we remove the possibility of the occurrence of a repulsive gravity
(negative $M$), then the bottom of the pit would be at $M_f =0$. At this
state, both $ E_g$ and $E_{in}$ would be infinite but  of opposite sign
and separated by a finite gap $M_0$ much like what happens in a
renormalized Quantum Field Theory.

There could be another confusion here as to how can $\mid E_g\mid$ be
infinite when $M_f=0$. This depends on how fast the value of $M \rightarrow
0$ with respect to $R\rightarrow 0$ and is perfectly allowed for a
singular state.

Nevertheless, it must be borne in mind that
 for an actual strong gravity case all such contributions shown in the
foregoing equation
intermingle with one another in a non-linear and inseperable manner.
\subsection {Principle of Equivalence (POE)}
Even though there are many published results suggesting $M=0$ in
connection with naked singularities, our work might be singled out with
the plea that a $M=0$ result violates POE. We repeat once again that, POE
only says that the local nongravitational laws of physics are the same as
the corresponding laws in STR. For example, this would mean that the
Stefan- Boltzman law which tells that the emissivity of a black body
surface is $\propto T^4$, remains unchanged. POE does not impose any limit
on the value of $T$ itself and hence on the total amount of radiation
emitted from the black body surface. POE does not say that only a certain
percentage of the initial total mass energy $M_i$ can be radiated in the
process, POE has got nothing to do with either the imposition of any
additional local constraint (such as a maximum value of $T$) or any global
issues.

If one would invoke POE to debar phenomenon which are not understandable
in Newtonian notions (like $M \equiv M_0$, incorrectly) GTR itself is to
be discarded. With such a viewpoint,  all work on Positive Energy Theorems
are to be considered as  redundant and unnecessary because in STR, the
mass-energy of a system which was positive to start with can never be
negative.

\subsection {Matter - Antimatter Annihilation ?}
In STR, there is no gravity and hence there is no Kelvin-
Helhmoltz process, neither could there be any real finite
material body held together by any long range force ( a plasma
has to be confined by external electromagnetic fields). And there
could be a naive idea that the entire initial mass energy may be
radiated only if there are processes like $e^+ e^- \rightarrow
2\gamma$. If this is envisaged as the only way to generate
radiation (in this case photons), it must be remembered that such
a thing refers to systems having total lepton number or total
baryon number as zero. For matter consisting of a definite baryon
number and lepton number there cannot be any energy extraction
by this process. Yet such matter radiates because of normal
electromagnetic processes like Bremsstralung or Compton
processes, or by nuclear processes like $ p+p
\rightarrow \pi^0 \rightarrow 2\gamma$. Actually at very high densities
and temperatures, in astrophysical scenarios, energy is liberated by the
so-called URCA or weak interaction processes involving emission of $\nu
\bar \nu$. Whatever be the process, if the global Kelvin- Helmholtz process
heats up the matter to sufficiently high temperature near the singularity
(to which everybody agrees), the center of mass energy of the colliding
particles, like, electrons, protons, neutrons, quarks or whatever it may
be, will be accordingly high enough.  And in this limit, for an individual
collision, the colliding particles can radiate not only an energy equal to
their rest mass but any amount higher than this. The easiest example would
be that an $e^- - e^+$ collider can generate particles (photons,
neutrinos, quarks etc.) much heavier than $0.5 MeV$. And it should be also
remembered that when we say that the entire $M_i c^2$ may be radiated, we
do not mean that this happens in a flash as is the case for
matter-antimatter annihilation.  On the other hand, in gravitational
collapse, it is the integrated radiation over the entire history of the
process we are concerned with.

\section { Dust Collapse}
There is no way we can ever think of exactly solving  even the adiabatic
collpase equations for a real fluid, i.e., one having pressure. Further,
this idea of adiabaticity would break down as soon as the fluid starts to
contract because of the Kelvin-Helmholtz energy liberation. Even if we
consider the fluid to be degenerate and at $T=0$ to start with,
gravitational contraction would keep on heating it up unless it acquires
an effective adiabatic index $\gamma_t =4/3$. On the other hand, we may
feign to ignore the role of any temperature in the fluid by artificially
assuming a polytropic EOS, $p\propto \rho^{\gamma_t}$ even when the gas is non-
degenerate. But the value of $\gamma_t$ will keep on evolving and it is not
possible to find any unique solution for the entire range of $p$ and
$\rho$ even by any numerical means. Depending on the inevitable
hidden assumptions
made, it may be possible to obtain various solutions
 and none of which
may have to do much with the actual complicated physics of atomic and
nuclear matter at arbitrary high density and pressure ($z_s=\infty$). And, the only way,
one may hope to obtain an exact or near exact solution, at the cost of the
actual thermodynamics, is to do away with the EOS, i,e., to set $p\equiv
0$ even when $\rho \rightarrow \infty$!! Even then, there could be {\em
exact} analytical solution only when the dust ball has uniform density.
And this problem was first (apparently) solved by Oppenheimer and Snyder (OS)\cite{4}.
It may be mentioned that recently the problem of the occurrence of the
final state in a GTR collapse, with the inclusion of pressure gradients,
have been considered by Cooperstock et al.\cite{8}. These authors have
correctly pointed out that at very high densities the effect of pressure
gradient must be included and for the likely occurrence of negative
pressure gradients, there may may not be BH formation at all.

 OS initially worked in the COF, but, then, to match the
internal solutions with the external ones, eventually shifted to the  (non
comoving) Sch. frame involving $R, T$. Without giving the details of the
actual mathematical manipulations, we shall simply present their key
equations. By matching the internal solutions with the exterior ones they
obtained a general form of the metric coefficients and also a relation
between $T$ and $R$ which is valid for the entire range of $ \infty > R
>0$ :
\begin{equation}
g_{TT} =e^\nu =  \left[ (dT/d\tau)^2 (1-U^2)\right]^{-1}
\end{equation}
\begin{equation}
-g_{RR} = e^\lambda =(1-U^2)^{-1}
\end{equation}
and,
\begin{equation}
T = {2\over3} R_{gb}^{-1/2} (r_b^{3/2} -R_{gb}^{3/2} y^{3/2}) -2 R_{gb} y^{1/2} +
R_{gb} \ln {y^{1/2} +1 \over y^{1/2} -1}
\end{equation}
where
\begin{equation}
y \equiv{1\over 2} \left[ (r/r_b)^2 -1\right] + {r_b R\over R_{gb}  r}
\end{equation}
It is the above Eq. (7.3) which corresponds to Eq. (36) in the OS paper.
OS also showed that the relation between $T$ and $\tau$ is determined by
\begin{equation}
F \tau + r^{3/2} = R^{3/2}
\end{equation}
where,
\begin{equation}
F = - (3/2) R_{gb}^{1/2} (r/r_b)^2 ; \qquad r\le r_b
\end{equation}
So, for the outer boundary, we have
\begin{equation}
\tau = {2\over 3} { r^{3/2} - R^{3/2} \over (r/r_b)^2 R_{gb}^{1/2}}
\end{equation}
According to OS, in the limit of large T, one can write
\begin{equation}
T \sim -R_{gb} ~\ln \left\{ {1\over 2}\left[ \left({r \over r_b}\right)^2 -3\right] + {r_b
\over R_{gb}} \left( 1- {3 R_{gb}^{1/2} \tau \over 2 r_b^2}\right)\right\}
\end{equation}
The last term of the above equation contains a typographical error, and,
in general, this equation missed a numerical factor of 4. The {\em corrected}
form should be
\begin{equation}
T \sim -R_{gb}~ \ln \left\{ {1\over 8}\left[ \left({r \over r_b}\right)^2 -3\right] + {r_b
\over 4 R_{gb}} \left( 1- {3 R_{gb}^{1/2} \tau \over 2 r_b^{3/2}}\right)\right\}
\end{equation}
From the foregoing equation, they concluded that, ``for a fixed value of $r$
as $T$ tends toward infinity, $\tau$ tends to a finite limit, which
increases with $r$''.

It follows  that the supposed finite limit for $\tau \propto R_{gb}^{-1/2}$.
OS then found the explicit expressions for the internal Sch.
metric coefficients in the limit of $T\rightarrow \infty$. However,
they missed a numerical factor of 4. And though this numerical factor does not
change the qualitative behavior of the solutions, in the following, we present the
{\em corrected} relevant expressions
\begin{equation}
e^{-\lambda} = 1 -(r/r_b)^2 \left\{4 e^{-T/R_{gb}} +{1\over 2} \left[3 - (r/r_b)^2\right]\right\}^{-1}
\end{equation}
and
\begin{equation}
e^\nu = {e^{\lambda- 2 T/R_{gb}}\over 4}
\left\{ e^{- T/R_{gb}} + {1\over 2} \left[3-(r/r_b)^2 \right]\right\}
\end{equation}
Note that these equations were obtained by eliminating $R$, and the $T=\infty$
limit covers the entire $R \le R_{gb}$ range.  Also note that the {\em
comoving coordinates} $r$ and $r_b$ are fixed, and when there is a
complete collapse to a physical point at $R=0$,  $e^\lambda$ {\em must
blow up irrespective of the value of $r$}. {\bf But this does not happen for
the OS solution for an internal point} $r<r_b$!

For instance {\em if} $R_{gb}>0$, for $r/r_b =0.5$, we obtain $e^\lambda =
15/13$ at $T=\infty$, when {\em we should have obtained} $e^\lambda
=\infty$. And for the central point $r=0$ (which also corresponds to
$R=0$), we find that after the collapse ($T=\infty$), we have
$e^\lambda=1$ when we should have had $e^\lambda =\infty$!  And OS,
somewhat casually, noted this: ``For $\lambda$ tends to a finite limit for
$R \le R_{gb}$ as $T$ approaches infinity, and for $R_b=R_{gb}$ tends to
infinity. Also for $R \le R_{gb}$, $\nu$ tends to minus infinity.''

However, unfortunately, OS did not bother to ponder on the
genesis of this completely unphysical aspect of their solution.
 This, on the other hand, is a
definite signature that {\bf there is a severe problem in the foundation of this
problem}.

\subsection{True Solution of the O-S Problem}
We note that  OS completely overlooked
 {\em the most important feature} of Eq. (7.3) (their Eq. 36),
that in view of the presence of  the $T \sim \ln {y^{1/2} +1\over y^{1/2} -1}$ term,
 in order that $T$ {\bf is definable at all}, one must have
$y \ge 1$

For an insight into the problem, we first focus attention on the outermost
layer where $y_b = R_b/R_{gb}$, so that the above condition becomes
\begin {equation}
{2GM_b/R_b c^2} \le 1
\end{equation}
Thus a careful analysis of the GTR homogeneous dust problem as enunciated
by OS themselves actually tell that trapped surfaces cannot be formed
even though one is free to chase the limit $R\rightarrow R_{gb}$.
This means that, the final gravitational mass of the configuration is
\begin{equation}
M_f (R=0) = M_g (R_b=R_{gb}) =0
\end{equation}
But then, for a dust or any adiabatically evolving fluid
$M_i = M_f = constant$.
Therefore, we must have $M_i = 0$ too. And {\em for a finite value of}
$R$, this is possible only if $\rho =0$. But for a dust $\rho =\rho_0 =
mn$ and therefore, we have $n=0$. Finally, the total number of the baryons in the configuration
$N=0$.
From, a purely mathematical view point,
the $N=0$ limit can be described as
\begin{equation}
r =r_b \rightarrow 0; \qquad r/r_b \rightarrow 1
\end{equation}

Further note  that
 if we really assume $R_{gb} \neq 0$, the second term $r_b
R/R_{gb} r$ of Eq.(7.4) can be made arbitrarily small as the
collapse proceeds to $R\rightarrow 0$. Remember here that $r$ and
$r_b$ are comoving coordinates and are fixed by definition. So
for any interior region $r$ separated from the boundary by a
finite amount $r <r_b$, $y$ {\bf becomes negative} in
contravention of Eq.(7.3) if $R_{gb} \neq 0$:
\begin{equation}
y \rightarrow {1\over 2} \left[ (r/r_b)^2 -1\right] <0 ~as ~ R\rightarrow 0
\end{equation}
 This is alleviated if
either or both of the two following conditions are satisfied : (i) $r=r_b$, as
derived above or (ii) $R_{gb} =0$, which again leads to the previous condition.

{\em Thus had O-S carefully noted this simple point, they would probably
not have
proceeded with the rest part of their paper which hints at the
formation of a finite mass BH in a completely erroneous manner}.
And mathematically, the Eqs. (7.10) and (7.11), in a self-consistent manner
 degenerate to  definite limiting forms:
\begin{equation}
e^{-\lambda} \approx 1 - \left(4 e^{-T/R_{gb}} +1\right) \rightarrow
0;\qquad e^\lambda \rightarrow \infty,
\end{equation}

\begin{equation}
e^\nu \approx {e^{\lambda - 2T/R_{gb}}\over 4} \left\{e^{-T/R_{gb}} +1\right\} \rightarrow 0
\end{equation}

\subsection {Common Perception About Formation of Horizon}
There is a widespread idea that atleast for a dust collapse, irrespective
of our above explicit proof, the formation of an Event Horizon is inevitable.
One assumes here a shell of dust particles to be either at rest or in
 equilibrium at $t=0$. Then {\em suddenly} the dust is
envisaged to lose its state of rest and is allowed to collapse.
And since a dust does not radiate its gravitational mass remains
fixed and it is expected to reach its horizon $R_{g} = 2GM/c^2$
in a finite proper time on the basis of the OS solution (which
ignored the $y\ge 1$ condition).

For a resolution of this puzzle, first one has to appreciate that if a
dust ball is ever at rest without the support of external mass energy, its
mass must be zero. To see this simply consider the Oppenheimer-Volkoff
equation\cite{37} for hydrostatic balance:
\begin{equation}
{dp\over dR}=- {p+\rho(0) \over R(R-2GM)}(4\pi p R^3 + 2G M)
\end{equation}
If $p=0$, the above equation yields $\rho=0$ too\cite{38} (Private Comm.,
A.K. Raychowdhury).
 From the view point of thermodynamics, any physically
meaningful EOS will yield $\rho =0$ if $p=0$ irrespective of whether the
fluid is in hydrostatic equilibrium or not.
And when $\rho=0$ everywhere, it can be shown that
the proper time for collapse to both the horizon and the central
singularity is $\infty$. In particular, the latter is given by
\begin{equation}
\tau_c = {\pi \over 2} \left({ 3\over 8\pi G \rho(0)}\right)^{1/2}
\end{equation}
where $\rho(0)$ is the density of the dust {\em when it was at rest}; $U(0)=0$.

For a finite value of $R_b(0)= R_0$, one has $M=0$ for $\rho =0$. And if
$R_0 =\infty$ to make $M >0$, one would again find the proper time to
reach the horizon to be $\infty$.
Now the reader is requested to carefully go through the following subtle
points for a  resolution of this puzzle.

$\bullet$~ One can imagine the dust ball to be in a state of rest by either
unconsciously assuming the presence of (1)  some finite pressure gradient forces
 or (2)  some external source of mass energy, like,
friction forces or biological forces at $t=0$. The latter
possibilty eventually boils down to the presence of some external
electromagnetic force field in the problem at $t=0$.

$\bullet$~ Thus at $t=0$ we do not have a dust ball, but, on the other
hand, we have either (1) a physical fluid or (2) a system of particles
interacting with some external (electromagnetic) sources of mass-energy.

$\bullet$~ If the fluid loses its hydrostatic balance, surely, it would
start collapsing at $t=0$, but {\em it does not mean that the collapse would be
a free fall}. In other words the relevant initial condition of the problem
that there is a ``sudden collapse'' can be realized, but, {\em what cannot be
realized is the idea that} $p =finite ~for~ t\le 0$ but again $p=0~
for~t\ge 0$. {\em This assumption  that pressure vanishes instantly violates
causality and is not allowed by GTR}.

$\bullet$~ In the latter case too, the external electromagnetic
field must {\em decay over a finite time scale, howsoever,
small}. And when this finite time scale removal of the external
electromagnetic field will be implemented, the problem would be
different from the ideal ``dust collpse''problem. The OS result
$\tau \propto M_i^{-1/2}$ {\bf will cease to be valid} in such a
case. And we must not let loose our intuitive Newtonian concepts
to determine the fate of such a problem by overrdiding the
general constraint $2GM/R \le 1$. In particular, here $R$ {\em
must not be confused as the proper length along the worldline}.

$\bullet$~A pure dust is however allowed to collapse from a state of rest
by starting from $\infty$. But once it statrs from $\infty$, it would never
reach the would be EH supposed to be located at a finite radial distance
in a finite comoving time. This would be in agreement with our contention.

\section{Kruskal Coordinates}
Since we have already shown in a most general manner that because
of an inherent {\em Global Constraint}, the Einstein equations
dictate that if the collapse proceeds upto $R=0$, the
gravitational mass of the singularity would be $M=0$, it is clear
that, if there would be any BH formed by gravitational collapse,
its mass would be zero. To show the consistency of our result, in
the following, we assume first the existence of a finite mass
Sch. BH, and then, try to verify whether a finite value of $M >0$
is allowed or not. Both the exterior and interior spacetime of a
BH is believed to be described by the Kruskal
coordinates\cite{39}. For the exterior region, we have (Sector
I and III):
\begin{equation}
u=f_1(R) \cosh
{T\over 4M}; \qquad v=f_1(R) \sinh
{T\over 4M}; ~~R\ge 2M
\end{equation}
where
\begin{equation}
f_1(R) = \pm \left({R\over 2M} -1\right)^{1/2} e^{R/4M}
\end{equation}
where the +ve sign refers to ``our universe'' and the -ve sign refers to the
``other universe'' implied by Kruskal diagram.
It would be profitable to note that
\begin{equation}
{df_1\over dR} = {\pm R\over 8M^2} \left({R\over 2M} -1\right)^{-1/2} e^{R/4M}
\end{equation}
And for the region interior to the horizon (Sector II and IV), we have
\begin{equation}
u=f_2(R) \sinh
{T\over 4M}; \qquad v=f_2(R) \cosh
{T\over 4M}; ~~R\le 2M
\end{equation}
where
\begin{equation}
f_2(R) = \pm \left(1- {R\over 2M}\right)^{1/2} e^{R/4M} =\sqrt{-1} f_1(R)
\end{equation}
where, again, the +ve sign refers to ``our universe'' and the -ve sign
refers to the ``other universe''.
It is found that,
\begin{equation}
{df_2\over dR} = {\mp R\over 8M^2} \left(1- {R\over 2M}\right)^{-1/2} e^{R/4M}
\end{equation}
Given our adopted signature of spacetime ($-2$), in terms of $u$ and $v$, the metric for the entire spacetime is
\begin{equation}
ds^2 = {32 M^3\over  R} e^{-R/2M} (dv^2 -d u^2) - R^2 (d\theta^2
+d\phi^2 \sin^2 \theta)
\end{equation}
The metric coefficients are regular everywhere except at the
intrinsic singularity $R=0$, as is expected. Note that, the
angular part of the metric remains unchanged by such
transformations and $R(u,v)$ continues to signal its intrinsic
spacelike nature. In either region we have
\begin{equation}
u^2-v^2= \left({R\over 2M} -1\right) e^{R/2M}
\end{equation}
so that
\begin{equation}
u^2-v^2 \rightarrow 0; \qquad u^2 \rightarrow v^2; \qquad R= 2M
\end{equation}
In the External region, we also have
\begin{equation}
{u\over v} = \coth (T/4M)
\end{equation}
For studying the Kruskal dynamics, it would be useful to briefly recall
the dynamics of a test particle in the Sch. coordinates because the
Kruskal coordinates are built by using $R$ and $T$.

\subsection{RADIAL GEODESIC IN SCHWARZSCHILD COORDINATE}
For a radial geodesic, we have
\begin{equation}
ds^2 =dT^2 (1- 2M/R) - (1-2M/R)^{-1} dR^2
\end{equation}

In terms of the conserved energy per unit rest mass
\begin{equation}
E = {dT\over d\tau} (1-2M/R)
\end{equation}
it is possible to find that\cite{17,18}
\begin{equation}
{dR\over dT} = -  {1-2M/R\over E} [E^2 - (1-2M/R)]^{1/2}
\end{equation}
If $E\rightarrow \infty$, {\em as it happens for a photon}, for arbitrary value
of $2M/R$, the foregoing equation attains a form
\begin{equation}
{dR\over dT} \rightarrow -  (1-2M/R)
\end{equation}
Interestingly, {\em even when $E$ is finite}, Eq. (8.13) {\bf attains  the same form}
if $R\rightarrow 2M$.

\subsection{Radial Kruskal Geodesic}
We would like
to {\em explicitly verify} whether the (radial) geodesics of
material particles are indeed {\em timelike} at the EH which they
must be if this idea of a finite mass Schwarzschild BH is
physically correct. First we focus attention on the region $R\ge
2M$ and differentiate Eq.(8.1) to see
\begin{equation}
{du\over dR} = {\partial u\over \partial R} + {\partial u\over \partial T}
{dT\over dR}= {df\over dR} \cosh {T\over 4M} + {f\over 4M} \sinh
{T\over 4M}{dT\over dR}
\end{equation}
Now by using Eqs. (8.1-3) in the above equation, we find that
\begin{equation}
{du\over dR} = {ru\over 8M^2} (R/2M -1)^{-1} + {v\over 4M} {dT\over dR}; \qquad
R\ge 2M
\end{equation}
Similarly, we also find that
\begin{equation}
{dv\over dR} = {rv\over 8M^2} (R/2M -1)^{-1} + {u\over 4M} {dT\over dR};
\qquad R\ge 2M
\end{equation}
By using Eq. (8.13) in the two foregoing equations and then by dividing
(8.16) by (8.17), we obtain
\begin{equation}
{du\over dv} = {u- {vE\over \sqrt{E^2 -1 +2M/R}}\over v-{uE\over \sqrt{E^2
-1 +2M/R}}}
\end{equation}
Interestingly, we have verified that one would obtain this same equation
for $du/dv$ even for Sectors II and IV ($R \le 2M$).
Since $u$ and $v$ are expected to be differentiable smooth continuous
functions everywhere except at $R=0$, and also since the ``other
universe'' is a mirror image of ``our universe'', we expect that the value
of $du/dv$ for any given $R$ must be the same, except for a probable
difference in the signature, in both the universes. For the ``our universe'',
it is found that, we have
\begin{equation}
u_H = v_H
\end{equation}
where ``$H$'' refers to the value on the event horizon $R=2M$. And it can
be found that if one approaches the horizon from Sectors II or IV (other universe),
one would have
\begin{equation}
u_H =-v_H
\end{equation}
In this case, we find
\begin{equation}
{du\over dv} = {u_H - v_H\over v_H -u_H} = {2 u_H\over 2v_H} = -1;~\qquad R=2M
\end{equation}
so that, we have $du^2 =dv^2$ at $R=2M$. Then, we promptly find
that for a radial geodesic
\begin{equation}
ds^2 = 16 M^2 e^{-1} (du^2-du^2)=0; ~~\qquad R=2M
\end{equation}
This implies that although the metric coefficients can be made to appear
regular, the radial geodesic of a {\em material particle becomes null} at
the event horizon of a finite mass BH in contravention of the basic
premises of GTR!

 And since, now, we cannot blame the coordinate system to
be faulty for this occurrence, the only way we can explain this result is
that {\em the Event Horizon itself corresponds to the physical
singularity} or, in other words, the mass of the Schwarzschild BHS
$M\equiv 0$. 

What would happen if we would approach the EH from sector I to III? In
this case, by taking appropriate limit, it can be shown that
\begin{equation}
{du\over dv} = {(e/u_H) - (u_H/ E^2)\over -(e/ u_H) + (u_H/E^2)}
\end{equation}
Here the magnitude of $du/dv$ appears to be a function of $E$ {\em if}
$u_H=v_H$ is finite. In particular, for $E=1$, again $du/dv = -1$
irrespective of the value of $u_H$ and again $ds^2=0$ on the EH.  Note
that, as long
as $ds^2 >0$ (time like), its value may be allowed to be a function of the initial
conditions, but once its value is 0 (null) for any initial condition, for the
sake of consistency, its value must be 0 for any other initial condition
too. This is similar to the following situation:
 As long as the value of
$V<1$, its value may be different for different observers and initial
conditions. But once any observer finds $V\rightarrow 1$, all other local
Lorentz observers would also find $V\rightarrow 1$. Coming back to the
case of $du/dv$, in order that $du/dv= -1$ at the EH for all the
observers, we must have $u_H = \infty$, which demands $M=0$.

And then, the entire conundrum of ``Schwarzschild
singularity'', ``swapping of spatial and temporal characters by $R$ and
$T$ inside the event horizon ({\em when the angular part of all metrics
suggest that $R$ has a spacelike character even within the horizon}),
``White Holes'' and ``Other Universes'' associated with the full Kruskal
diagram get resolved.

\subsection{Physical speed at the Horizon}
As mentioned earlier, the physical velocity of a particle under the
influence of a {\em static} gravitation field, as measured by a certain
local observer can be found by using Eq.(88.10) or the the preceding
unnumbered lowermost equation of pp.  250 of Landau and Lifshitz\cite{15}:
\begin{equation}
V = {dl\over d\tau} = {\sqrt{-g_{11}} dx_1\over \sqrt{g_{00}} dx_0}
\end{equation}
Thus for the Kruskal case, the radial speed of free fall
 is
\begin{equation}
V_K = {du\over dv}
\end{equation}
While extending this idea of a ``locally measured'' 3-speed one point is
to be borne in mind. As emphasized by Landau and Lifshitz\cite{15}, 
 there cannot be any ``static observer'' at the
Event Horizon. However, we
can conceive that there is a static observer at $R= 2M +\epsilon$, where
$\epsilon \rightarrow 0$. In other words, we may only study the limiting
behaviour of $V$ at the EH.
Thus, our result shows that the free fall speed at the EH, $V \rightarrow 1$, and
this is not allowed by GTR unless $R_{gb}=2M=0$. Now we explain why $V_{K}
\rightarrow 1$ at the EH for any coordinate system, Kruskal or Lemaitre or anything
else. Let the speed of the static other observer be $V_{Sch-O}$ with
respect to the Schwarzschild observer. By principle of equivalence, we can
invoke special theory of relativity locally. Then the free fall speed of
the material particle with respect to the other static observer will be
\begin{equation}
V = {V_{Sch} \pm V_{Sch -O} \over 1 \pm V_{Sch} V_{Sch-O}}
\end{equation}
We also find that for the Sch. metric\cite{17,18}, we have
\begin{equation}
V_{Sch} = \left[1- {1\over E^2}\left(1- {2M\over r}\right)\right]
\end{equation}
Note that for a photon, $E=\infty$, and the above Eq. correctly yields
$V_{photon} =1$ anywhere. And for a material particle having $E >0$,
it shows that, again $V_{Sch} \rightarrow 1$ as $R\rightarrow 2M$.
 Consequently, at the Event Horizon, Eq.(8.25)  would always yield $\mid
V\mid \rightarrow 1$ too.
{\em The value of $V$ can change in various coordinates only as
long as it is subluminous to all observers}.

\subsection{Back to Schwarzschild Coordinate System}
We found that $ds^2(u,v) =0$ at $R=2M$. But actually $ds^2$ is invariant
under coordinate transformation, and, there it should be obvious that
$ds^2(R,T) =0$ too at $R=2M$. And it is easy to verify that it is indeed so.
We found in Eq.(8.14) that {\em either for a photon  anywhere or for a material
particle at} $R\rightarrow 2M$,
\begin{equation}
(dR/dT)^2 \rightarrow (1-2M/R)^2
\end{equation}
Infact this above condition is called ``null geodesic'' one, and we find
that, when this is satisfied, we have
\begin{equation}
 {dR^2\over 1-2M/R} \rightarrow   (1-2M/R)dT^2= dz^2
\end{equation}
By using this Eq. in Eq.(8.11), we find that
 for radial geodesics, {\em either for a photon anywhere or for a material particle at}
$R=2M$, we have
\begin{equation}
ds^2 (R,T)\rightarrow dz^2 -dz^2 =0
\end{equation}
Normally one would ignore this result as a reflection of the`` coordinate
singularity''. But, even then, the value of $ds^2$ is an invariant, and
its value must be same in all coordinates, as we have already verified for
the Event Horizon.
Sadly, like a proverbial ``darkness beneath the lamp'' this simple point
was not seriously taken note of in the past!

\section{Summary and Conclusions}
We found that for the
continued collapse of any perfect fluid possessing arbitrary EOS and
radiation transport properties, a proper amalgamation of the inherent
global constraints arising because of the dependence of spatial curvature
like parameter ($\Gamma$) on the global mass-energy content, $M$,
 directly shows that {\em no trapped surface is
allowed by GTR}. This result becomes independent of the details of the
radiation transport properties because the integration of the (0,0)-
component of the Einstein equation, yields the definition of $M$ by
absorbing all  quantities like $\rho$,  $q$, and $H$, in whichever fashion they
may be present. Then it follows that if there is a continued collapse,
the final gravitational mass of the
configuration necessarily becomes zero.  This $M_f=0$ state must not be
confused as a vacuum state, on the other hand, the baryons and leptons are
crushed to the singularity with an infinite negative gravitational energy $E_g
\rightarrow -\infty$. On the other hand, the
positive internal energy is also infinite $E_{in} \rightarrow
\infty$.
However, in the present paper, we did not investigate whether
this state corresponds to $2GM_f/R <1$ or $2GM_f/R =1$. In another
work\cite{40},
 we find that, it is the latter limit which should be appropriate, i.e.,
the system keeps on radiating and tends to  attain the
state of a zero mass  BH characterized by zero energy and entropy, the
ultimate ground state of classical physics. Since the concept of a BH is
intrinically meaningful only for $M >0$, by borrowing a terminology from
Cooperstock et al.\cite{8}, we may call this final state to be {\em
marginally naked}. In fact the Eq. (17) of this paper\cite{8} considered
the possibility that as $R\rightarrow 0$, one may have $M\rightarrow 2M
\rightarrow 0$. And it is precisely this result which we have obtained
without making any kind of assumption.

In this paper, we did not try  to find
 the proper time required to attain this absolute classical singular
ground state though we found that for the fictitious dust solutions $\tau =\infty$.
This question has, however, been explored elsewhere\cite{40} to find that, for a
real fluid too, $\tau =\infty$. This means that there is no incompleteness
in the radial worldlines of the collapsing fluid particles inspite of $R$
having a finite range ($r_b$). Such a Non-Newtonian behavior is
understandable in GTR because, it was found\cite{40} that although
$M$ keeps on decreasing, the curvature components $\sim GM/R^3 \sim
R^{-2}$ and $-g_{rr}$ tend to
blow up. As a result the 3-space gets stretched and stretched by the strong
grip of gravity, or in other words, the proper distances eventually tend to blow up too.

We also found that this inherent global constraint is also imprinted in the
 important work of Oppenheimer and Snyder\cite{4} because
in order that, at the boundary of the star,
\begin{equation}
T \sim \ln {y^{1/2} +1
\over y^{1/2} -1} =\ln {(R_b/R_{gb})^{1/2} +1
\over (R_b/R_{gb})^{1/2} -1}
\end{equation}
 remains {\bf definable}, one
must have $2GM_b/R_b\le 1$. And then the central singulaity $R_b \rightarrow
0$ could be reached only if $M_b=0$, if the horizon coincides with the
central singularity.
 Accordingly, the value of $\tau_{gb} = \infty$
along with $T=\infty$. In fact this result {\em follows in a trivial fashion}
from Eq. (32) of their paper (our Eq.[8.10])
\begin{equation}
y\equiv {1\over 2} [(r/r_b)^2 -1] + {r_b \over r} {R\over R_{gb}}
\end{equation}
where the parameter $y$
which {\bf must be positive}. But if $R_{gb} \neq 0$,
 as $R\rightarrow 0$, it is trivial to see that $y$
 {\bf actually becomes negative} for $r<r_b$. 
This shows that actually the horizon or any trapped
surface in never allowed by the OS solution.
And we noted that since OS did not incoroprate this intrinsic constraint
in their eventual approximate expression for $e^\lambda$, {\bf it failed to
blow up for the internal regions of the dust ball even when the collapse
is complete}.

Despite having proved our results in a general fashion, we reconsidered
the case of a finite mass Sch. BH described by Kruskal coordinates. We
found that the radial geodesic of a material particle, which must be
timelike at $R=2M$, if indeed $M>0$, actually becomes null. And this
independently points out that $M=0$.
This simple fact independently asserts that
 there is no Event Horizon, no  Schwarzschild Singularity, no T-region,
 and the only singularity that might have been present is the central singularity,
and  whose mass must be zero. Technically, one might view this
central singularity as the Schwarzschild Singularity associated with a
zero mass BH. Even then the existence of such a zero mass BH could be
realized only if the collapse process could be complete in a finite proper
time; but it actually takes infinite time : Nature abhors not only naked
singularities but all singularities; and we find that only GTR may be having the
mechanism of removing such singularities even at a classical level. This
precise possibilty has recently been considered by Senovilla\cite{29}. And
this happens because of the marriage between the physics and (spacetime)
geometry. If somehow, one would try to
build up a super concentrated energy density
near a ``point'', the space would get dynamically stretched by the gravity associated with
the concentrated energy density  and a singularity is avoided. But we
emphasize that the present discussion does not rule out the likely occurrence of
a singularity in the cosmological context.

Consequently, several associated theoretical
confusions like (i) whether the physically defined circumference coordinate $R$
can, suddenly become a time-like coordinate, (ii) whether there could be White
Holes freely spewing out matter and energy in the observable universe,
 and (iii) whether information can really be lost from
the observable universe in violation of the quantum mechanics, which have
plagued GTR in the present century, would be resolved, if the present work is
correct.

Finally, we appreciate the physical intuition of Einstein\cite{42} and
Landau\cite{43}  in
not being able to accept the reality of Schrawzschild Singularity
or any singularity in GTR. We also recall that  Rosen\cite{44}, in an
unambiguous manner noted the impossible and unphysical nature of the T-region.

``{\em so that in this region $R$ is timelike and $T$ is spacelike. However,
this is an impossible situation, for we have seen that $R$ is defined
in terms of the circumference of a circle so that $R$ is spacelike, and we
are therefore faced with a contradiction. We must conclude that the
portion of space corresponding to $R <2M$ is non-physical. This is a situation
which a coordinate transformation even one which removes a singularity can
not change. What it means is that the surface $R=2M$ represents the
boundary of physical space and should be regarded as an impenetrable
barrier for particles and light rays}.'' (Emp. by author).

We have, in this paper, attempted to resolve all such paradoxes by showing that not only
the $R <2M$ region unphysical, it does not exist or is not ever created.

We have pointed out several instances when it was hinted or
suggested that the GTR singularity may correspond to $M=0$ state.
Additionally we point out that the numerical studies of collapse
of scalar fields suggest that it is possible to have BHs of
$M=0$\cite{44}. More importantly, the supersymmetric string
theories find the existence of extremal BHs with charge $Q=M$,
which for the chargeless case yields $M=0$\cite{45}. However,
ironically, string theorists, at this moment, are guided by the
erroneous notion that GTR yields BH with $M>0$, and are
struggling to wiggle out of this result by modifying the definition
of event horizon into the socalled ``stretched horizon''.

Although, it might appear that astrophysics would be poorer in
the absence of the mystique of BH, actually, it may be possible
to envisage new varieties of stable or quasi- stable ultracompact
compact objects of stellar mass or dynamically contracting super
massive stars responsible for new gamut of astrophysical
phenomenon. But there is one
constraint imposed by GTR on such probable {\em static} 
ultracompact objects :  if the compact object is assumed to be {\em cold}
and in hydrostatic equilibrium, the surface redshift $z_s =
\Gamma^{-1} -1 <2$\cite{16}, and it does not mean that there can
not be any compact object beyond this limit, i.e., $z_s \ge 2$.
It only means that such high $z_s>2$ objects must be ``hot'' and
dynamically contracting (remember the time to collapse to a
singularity is $\infty$). Also, in case there is a positive (repulsive)
cosmological constant, it may be possible to have more massive static
compact objects.

Also recall here that if rotation is taken into considerations, the value
of $M_{OV}$ could be significantly raised. But to fully appreciate the
question of likely existence of stellar mass BH candidates of masses as
high as $\sim 10 M_\odot$, we must keep an open mind with regard to our
present day understanding of QCD. Even with reference to present state of
knowledge of QCD, there could be compact objects with exotic EOS, where
the masses could be $\sim 10 M_\odot$ or even higher\cite{46}. These stars are
called Q-stars (not the usual quark stars), and they could be much more compact than
a canonical NS; for instance,
a stable  non rotating Q-star of mass 12$ M_\odot$ might have a radius of
$\sim 52$ Km. This may be compared with the value of $R_{gb} \approx
36$ Km of a supposed BH of same mass.

In general, it is believed that, at sufficient high temperature, quark
confinement may melt away. And the energy gained from the pairing of quarks
and antiquarks of all colors which drive the chiral symmetry breaking may
be overcome by the entropic advantage in letting the particles be free. At
a very high $T$, therefore, asymptotically, free quarks, antiquarks and
gluons should be liberated\cite{47} and provide new sources of pressure. There is already
some
 evidences that at a temperature of $\sim 150$ MeV,
there is  a phase transition in hot nuclear matter and new degrees of
freedom are suddenly liberated. It is such processes which may allow
ultracompact objects to be in a stable or dynamic quasi-stable state.

However, the above argument that there there may be {\em static}
ultracompact objects having masses larger than canonical neutron stars
does not at all mean that we are advocating that all the central compact
objects hinted in X-ray binaries or Active Galaxies must be static. On the
other hand, our work shows that there could be compact objects which would
take {\em infinite proper time} to collapse a singular state because
although, in the coordinate space the objects may be sinking, in the inner
physical space (proper radius), they may be expanding! In other words, 
the final state suggested by our work (for a physical fluid) is the one with
$\Gamma \rightarrow 0$, $U\rightarrow 0$, but with $V \rightarrow c$.
Recall, when we say ``radius'' of a body we imply $R$, and thus the rate of observed
radial contraction in such cases would be given by $dR/dT \rightarrow 0$. Yet,
in internal space, the body may be contracting with a speed $V \rightarrow c$ !
 But,
at a
finite proper and lab time, when we detect a massive condensation or ECO,
it is likely to have a value of $R$ almost equal to its
Schwarzschild radius and the value of $g_{00}$ on the surface could be
exremely low {\em but finite}. For example  it may have a value of
$g_{00} =10^{-10}$, $10^{-11}$ or even less (but finite).
Since, the corresponding lab speed of
contraction $dR/dT \approx 0$, the ECO   would look like a static and
frozen object over time scale of years. Yet, its actual local speed of
collapse (in the space of proper radial distance) $V \sim (g_{00})^{-1}~
dR/dT\approx c $ or also $V\ll c$.
 We understand that it is extremely
difficult to comprehend this because, intuitively, we tend to equate $R$
as the measure of proper radial distance. 

When
advection dominated accretion flow will interact with such an Eternally
Collapsing Object (ECO), in case, indeed, $V \approx c$, the {\em
collision process would emit insignificant radiation}. However, if one
assumes that the central object is {\em static}, one would expect large
luminosities from the surface of the object. And the absence of such
luminosities (in case $V \approx c$) would be interpreted as the
``evidence for an event horizon''! An ECO is struggling to attain the
$R=0$, $g_{00} =0$ and $g_{rr} =\infty$ state, but it would succeed to reach
there only after the elapse of infinite proper time by which time its
gravitational mass $M\rightarrow 0$.

Let us briefly recall the case of the recently discovered
 Cosmological Gamma Bursts like GRB990123\cite{48} and GRB980329\cite{49} which
 have been estimated to have radiated equivalent of $Q_\gamma \sim 2 M_\odot
c^2$ in the electromagnetic band alone under assumption of isotropy. The
total energy radiated including the neutrino emission is expected to be
atleast twice this amount  $Q \sim 4 M_\odot
c^2$. And this is in agreement with the predicted and estimated true
energy budget of poweful Gamma Ray Bursts\cite{50}. It is possible that
many GRB afterglows are highly beamed and the actual energy budget could
be considerably lower than estimated assuming isotropy. Note, while a BH cannot
have any intrinsic magnetic field, an UCO/ECO could be highly magnetized
and thus the latter is much more capable to explain likely beamed emission. 
In fact,
 the observed spectral break and rapid fading of the optical afterglow\cite{51}
in GRB990123 has been interpreted in terms of beaming. But this
interpretation is not fully satisfying because (i) the spectral break 
may be  explained by spherical models\cite{52} and (ii) no linear
polarization has been detected in the optical afterglow\cite{53}. In
general, the long term afterglow observations of GRB 970508 and
971214, show that they are inconsistent with jet models\cite{54}. Further
the afterglow of GRB 970402, 970616 and 98042588 too are consistent with
isotropic models\cite{55}. Such energy emission can be explained as
collapse of massive stellar cores, and is hardly possible if trapped
surfaces really formed at values of $M_f \approx M_i$. On the other hand
such phenomenon might be signalling the formation of new relativistic
ultracompact objects or ECOs.
\vskip 0.5cm
Acknowledgement: I benefitted from multiple reports of some eight
referees. I want to thank all of them but especially the referees whose
patient and constructive criticism has resulted in a consensual version of
my paper.

%\newpage

%{\Large \bf {References}}
%\begin{itemize}

\end{document}